\documentclass[a4paper]{spie}  

\usepackage{amsmath,amsfonts,amssymb}
\usepackage{graphicx}
\usepackage[colorlinks=true, allcolors=blue]{hyperref}
\usepackage{color}

\title{A snapshot full-Stokes spectropolarimeter\\for detecting life on Earth}

\author[a]{Frans Snik}
\author[a]{Christoph U.~Keller}
\author[a]{David S.~Doelman}
\author[b]{Jonas K\"uhn}
\author[c]{C.~H.~Lucas Patty}
\author[b,d]{H.~Jens Hoeijmakers}
\author[e]{Vidhya Pallichadath}
\author[e]{Daphne M.~ Stam}
\author[b]{Antoine Pommerol}
\author[f]{Olivier Poch}
\author[b]{Brice-Olivier Demory}

\affil[a]{Leiden Observatory, Leiden University, P.O. Box 9513, 2300 RA Leiden, The Netherlands}
\affil[b]{Physikalisches Institut, Universit\"at Bern, Sidlerstrasse 5, 3012, Bern, Switzerland}
\affil[c]{Biological Research Centre, Hungarian Academy of Sciences, P.O. Box 521, H-6701 Szeged, Hungary}
\affil[d]{Observatoire de Gen\`eve, Universit\'e de Gen\`eve, Chemin des Maillettes 51, 1290 Versoix, Switzerland}
\affil[e]{Faculty of Aerospace Engineering, Delft University of Technology, Kluyverweg 1, 2629 HS Delft, The Netherlands}
\affil[f]{Univ. Grenoble Alpes, CNRS, IPAG, 38000 Grenoble, France}

\authorinfo{Further author information: Frans Snik: E-mail: \href{snik@strw.leidenuniv.nl}{snik@strw.leidenuniv.nl}}

\begin{document} 
\maketitle

\begin{abstract}
We present the design of a point-and-shoot non-imaging full-Stokes 
spectropolarimeter dedicated to detecting life on Earth from an orbiting platform like the ISS. 
We specifically aim to map circular polarization in the spectral features of chorophyll and other biopigments for our planet as a whole. 
These non-zero circular polarization signatures are caused by homochirality of the molecular and supramolecular configurations of organic matter, and are considered the most unambiguous biomarker. 
To achieve a fully solid-state snapshot design, we implement a novel spatial modulation that completely separates the circular and linear polarization channels. 
The polarization modulator consists of a patterned liquid-crystal quarter-wave plate inside the spectrograph slit, which also constitutes the first optical element of the instrument. 
This configuration eliminates cross-talk between linear and circular polarization, which is crucial because linear polarization signals are generally much stronger than the circular polarization signals. 
This leads to a quite unorthodox optical concept for the spectrograph, in which the object and the pupil are switched. 
We discuss the general design requirements and trade-offs of 
LSDpol (Life Signature Detection polarimeter), a prototype instrument that 
is currently under development.
\end{abstract}


\section{INTRODUCTION}

Discovering extraterrestrial life is one of the greatest and most exciting scientific challenges for humanity in the 21st century. 
The exploration of our Solar System has led to the identification of potential habitable environments which might host life locally, in the sub-surfaces of Mars or in the liquid water reservoirs of some of the icy moons of the giant planets (Europa, Enceladus). 
Moreover, we are very much in the middle of the exoplanet revolution, with almost daily detections of planets orbiting other stars than the Sun.
Over the last decades, the mostly indirect exoplanet detection techniques 
that involve measurements of starlight (e.g., radial velocity and transit methods) 
have unveiled the spectacular ubiquity and diversity of exoplanetary systems. 
It is now clear that rocky planets in the habitable zones of stars 
(solar-type down to M-dwarfs) are indeed very common.
The next generation of telescopes, both on the ground and in space, will be 
optimized and equipped with advanced instrumentation for the direct observation 
and characterization of such potentially life-bearing planets.
Through spectroscopy and polarimetry, evidence for habitability and 
even habitation can be accrued\cite{Schwieterman2018}: 
Liquid water can be identified, for example,~through the glint of starlight off a surface ocean \cite{Trees2019}, or through the occurrence of rainbows that are produced by liquid-water clouds.
We also need to find indications of the planetary atmosphere being out of chemical equilibrium, e.g.~large amounts of molecular oxygen, which on Earth is a strong biomarker, as virtually all this oxygen is produced by photosynthesis.
However, also abiotic processes, such as UV photolysis from stellar flares, can produce O$_2$, and therefore a wide range of biomarker gases and atmospheric 
properties (in particular clouds) need to be detected and characterized 
before a detection of life can be robustly claimed.

One property of life is considered to be a universal and unambiguous biomarker: 
the \emph{homochirality} of molecules that constitute the building blocks of life (e.g.~amino acids, sugars) \cite{Wald1957,Bonner1995}.
The occurrence of a predominance of single handedness of chiral molecules is 
a clear sign of biological production, as pure chemistry would produce a 
racemic mixture of both left- and right-handed versions.
Multiple scattering of UV light in the early solar system is hypothesized to contribute to an initial enantiomeric excess that led to life on Earth picking left-handed amino acids and right-handed sugars\cite{Bailey+1998}.
The symmetry-breaking in chiral handedness leaves a distinct imprint on 
starlight that is reflected/scattered off organisms built up from 
homochiral molecules: 
it becomes partially circularly polarized. Because the direction
of circular polarization is determined by the chiral molecules,
it is not averaged out when this reflected/scattered light signal is
integrated across the planetary disk. This averaging out does
occur for the circular polarization signal of
starlight that is scattered by cloud and/or aerosol particles 
\cite{RossiStam2018}.
Furthermore, this circular polarization generally has a specific 
spectral signature with both negative and positive circular polarization peaks, 
which makes it readily distinguishable from multiple-scattering effects that can 
create spectrally flat non-zero circular polarization\cite{RossiStam2018}.

On Earth, particularly photosynthetic organisms exhibit a non-zero circular 
polarization spectrum, due to the homochirality of chlorophyll and other 
biopigments, see Ref.\cite{Patty+2019c}~and references therein.
Sparks et al.\cite{Sparks+2009a,Sparks+2009b}~describe laboratory experiments 
using a scanning monochromator with a photoelastic modulator (PEM) system to 
measure the circular polarization spectra due to transmission and reflection 
of various types of leaves and microbes, and show that the signals are weak
($\sim$1--0.01\%) but very pronounced.
Sterzik \& Bagnulo, used a spectropolarimeter on a telescope to measure the
spectral circular polarization signature of a Philondendron 
leaf\cite{SterzikBagnulo2011}.
In recent years, Patty et al.\cite{Patty+2017,Patty+2018,Patty+2019a,Patty+2019b}~have developed and utilized the TreePol instrument to enable sensitive circular spectropolarimetric measurements both in the lab and in the field. TreePol is spectrally multiplexed with fast, dual-beam ferro-electric liquid-crystal (FLC) modulation.
They demonstrated in the lab that the typical antisymmetric circular polarization signal in the chlorophyll absorption band for a leaf in transmission disappears as a leaf ``dies'', which they attribute to the desintegration of the chiral macromolecular organization of the chlorophyll molecules as opposed to the destruction of the chlorophyll itself\cite{Patty+2017}. Outside the lab, 
they used TreePol to obtain a clear non-detection of circular polarization for a nearby field of artificial grass, while a forest at $\sim$3 km distance yielded a clear remote detection of life\cite{Patty+2019b}.

We are currently building a second version of TreePol to fly on an aircraft 
to provide more observations of a larger and more varied range of scenes.
The TreePol design, however, cannot be adopted for a space mission for reasons
explained in Sect.~\ref{modulation}. We therefore started the design of a 
radical optical
concept that we call LSDpol, for ``life signature detection polarimeter'' 
(although it also alludes to a few crazy design choices made in the process).
The goal of this work is to design and build an instrument that can detect and 
map circular spectropolarimetric signatures of photosynthetic life on Earth in 
a large field survey, in an aircraft campaign, and in a global-mapping space
mission.
Our final aim with LSDpol is to observe the Earth as an exoplanet, to demonstrate
that the average circular polarization signal for planet Earth is strong enough to be detected when all the sunlight it reflects is compressed into a ``pale blue dot''. 

With large ground-based telescopes, the linearly polarized signatures of atmospheric molecular  oxygen and the ``green bump'' and ``red edge'' due to vegetation have been detected in Earthshine data \cite{Sterzik+2012,Sterzik+2019,Takahashi+2013,Bazzon+2013,MilesPaez+2014} (i.e.\ sunlight that has been reflected by the Earth and then by the Moon).
However, the weak circular polarization signatures are very difficult to measure in this way, as the depolarization of the lunar surface is significant and 
unknown, and, as the linear polarization is typically orders of magnitude larger, the polarization cross-talk issues in telescopes and instruments inhibit robust detections\cite{2011ASPC..449...76B}.
Moreover, Earthshine observations can only be done for a limited phase
angle range of the Earth.
Hence, a Dutch team is developing the LOUPE instrument (Lunar Observatory for Unresolved Polarimetry of Earth)\cite{Karalidi+2012,Hoeijmakers+2016} to provide highly time-resolved linear spectropolarimetry for all phase angles as seen 
from the Moon (i.e.~the observations should cover at least roughly a month).
LOUPE does not (yet) include the highly dedicated observation of circular
polarization as it greatly complicates the instrument design.

For the LSDpol instrument, we focus on mapping and monitoring the circular
polarization spectral signatures across the Earth first from a Low Earth Orbit, 
e.g.~from the ISS or a cubesat, and then possibly from a geo-stationary orbit or indeed the Moon. 
Because it is compact and robust (no moving parts), our novel instrument 
concept could also be used to detect extraterrestrial life, e.g.~on 
Mars\cite{SparksMars} as observed from an orbiter or lander/rover, or 
on icy moons like Europa and Enceladus with a mission probing the
geysers of sub-surface oceans.
In the end, we aim to inform the design of high-contrast imaging instrumentation\cite{HCIreview1,HCIreview2,HCIreview3} on the next generations of telescopes on the ground (the Extremely Large Telescopes targeting 
potentially habitable rocky planets orbiting M-dwarfs) and in space 
(targeting Earth analogs around sun-like stars), to ensure that their 
discoveries of signs of life on exoplanets are as unambiguous and
compelling as possible.

\section{INSTRUMENT REQUIREMENTS}\label{req}

For our LSDpol development we aim for a compact instrument with no moving 
parts that can be installed within the new Bartolomeo\footnote{\href{https://directory.eoportal.org/web/eoportal/satellite-missions/i/iss-bartolomeo}{https://directory.eoportal.org/web/eoportal/satellite-missions/i/iss-bartolomeo}} platform of the ISS.
This platform offers near-nadir observations of a large fraction of the 
Earth with a wide range of ecosystems (e.g.~rain forests, algal blooms, 
deserts, and ice/snow caps), in combination with ample data bandwidth.
We have imposed the following requirements on LSDpol:
\begin{itemize}
    \item Single spatial pixel (i.e.\ no spatial resolution) from ISS;
    \item Full-Stokes (i.e.\ Stokes parameters $I$, $Q$, $U$, and $V$~);
    \item Spectral range 350--900 nm;
    \item Spectral resolution $<$ 10 nm;
    \item Systematic error in circular polarization $<$ 10$^{-4}$ of the intensity.
    \item Cross-talk from linear polarization into circular $<$ 10$^{-3}$;
    \item Able to integrate over many observations to approximate polarized spectrum of exoplanet Earth;
    \item No moving parts;
    \item Insensitive to temperature (given a reasonable operational range); 
    \item Volume: $\sim$1 liter;
    \item Weight: $\sim$1 kg;
    \item Power: $\sim$1 W.
\end{itemize}
Before a full science mission on the ISS, we aim to deploy a prototype 
in the field and on an aircraft.
Moreover, the design should also be compatible with a cubesat mission 
(with more limited data bandwidth).

\section{POLARIZATION MODULATION APPROACH}\label{modulation}
The greatest challenge for LSDpol are the requirements on the polarimetric 
performance\cite{SnikKellerreview, Snikreview}, and those thus drive 
its entire design.
The TreePol instrument combines several polarization modulation techniques 
to ensure high sensitivity for circular polarization whilst being insensitive 
to linear polarization signals, which are usually several orders of magnitude 
larger\cite{Patty+2017}.
TreePol features a Fresnel rhomb that turns circular polarization into linear 
polarization at $\pm$45$^{\circ}$, which is consecutively measured by a rapidly 
switching FLC that is synchronized with the spectrograph detectors.
The polarization analyzer is a polarizing (wire-grid) beam-splitter that feeds 
two fiber-based spectrometers that are also synced to each other.
In this way, the two circular polarization states are modulated in time as 
rapidly as possible to minimize the influence of time-variable effects that 
could mimic actual variation in polarization signals (e.g.~swaying trees).
The dual-beam implementation adds another layer of mitigation of temporal
systematics, and also cancels the differential effects between the two 
spectrograph channels to first order\cite{Bagnulo+2009}.
In addition, a rotating half-wave plate in front of the instrument 
effectively depolarizes linear polarization while it merely flips the sign 
of circular polarization. 
This eliminates linear$\rightarrow$circular polarization cross-talk 
particularly by non-ideal effects of the FLC (e.g.~imperfect switching angle,
polarized spectral fringes).

Unfortunately, we consider this TreePol implementation to be 
generally unsuitable for space applications.
First, although a different kind of liquid-crystal modulation has recently been 
space-qualified\cite{GarciaParejo:19}, having an actively switching liquid-crystal element as the primary polarization modulator adds to the risk of instrument failure.
Moreover, in space, a fast rotating half-wave plate should be compensated by a counter-rotating element, which would further complicate the instrument. 
Furthermore, to assess systematic polarimetric errors as accurately as possible, 
we wish to also measure the linear polarization.
But most importantly, we opt for a fully snapshot polarimetric implementation
(thus, an instantaneous measurement of all Stokes parameters), 
as temporal modulation in combination with an orbiting, and possibly 
scanning platform could lead to spurious polarization signals. 

We have therefore investigated various implementations of ``channeled 
spectropolarimetry''\cite{Kudenovchapter,Snikreview}.
The most common implementation uses two thick retarder crystals with different 
thicknesses (often 1:2) that yield three carrier waves that spectrally 
modulate $Q$, $U$ and $V$\cite{OkaKato1999}.
However, $U$ and $V$ are modulated on the same carriers, only with a different
phase. Even a slight calibration error in the linear polarization $U$ can 
therefore obfuscate the much smaller circular polarization signal $V$.
A dual-beam implementation is necessary to prevent aliasing between the intensity
spectrum and the spectral polarization modulation carriers\cite{Snik+2009}.
Moreover, the spectral resolution of the polarization data product is necessarily
much smaller than the intrinsic resolving power of the spectrograph.
Sparks et al.\cite{Sparks+2012} introduced a similar implementation but 
with wedged birefringent prisms that provide modulation of all polarized Stokes 
parameters mostly in the dimension along the spectrograph slit.
This ensures that the full resolution of the spectrometer can be attained, 
and many photons can be collected in one shot on a two-dimensional detector, 
which facilitates a high polarimetric sensitivity.
However, $U$ and $V$ are still situated on the same carriers, which 
greatly enhances the risk of cross-talk problems.
Moreover, the authors reformat light from a target by means of cylindrical
fore-optics, which are also likely to induce cross-talk and instrumental 
polarization.

\begin{figure} [!t]
   \begin{center}
   \includegraphics[width=0.85\textwidth]{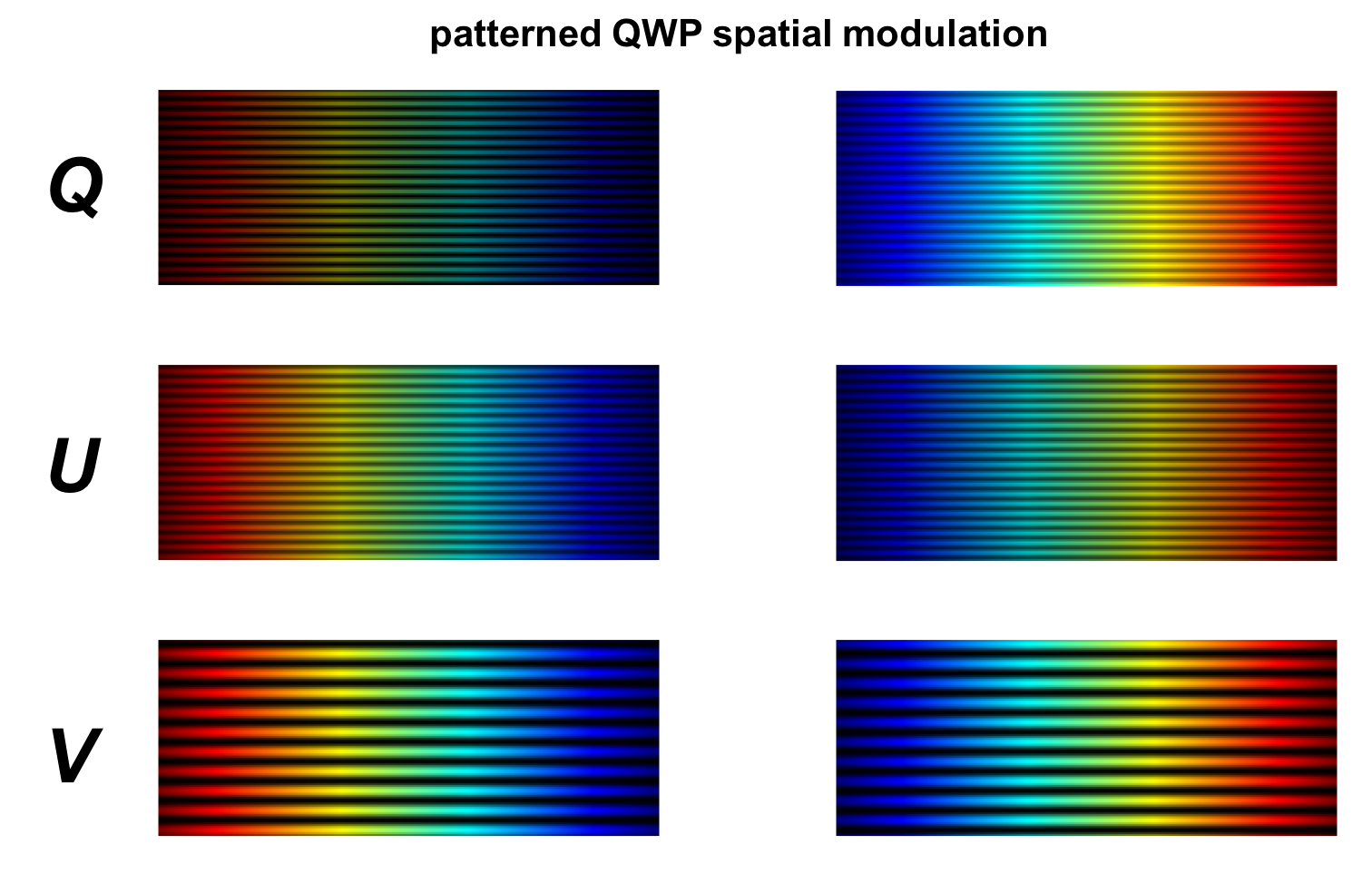}
   \end{center}
   \caption{\label{fig:polmod1} 
            Dual-beam (linear polarization splitting) implementation 
            of spatial modulation by a patterned quarter-wave retarder 
            at the slit for 100\% $Q$, $U$, and $V$, 
            see Eq.~\ref{rotret}.
            }
\end{figure}

A new solution to full-Stokes modulation was developed independently 
by Sparks et al.\cite{Sparks+2019} and the authors of this paper.
The implementation is very similar to a classical rotating wave plate 
polarimeter that yields a modulation cf.:
\begin{equation}\label{rotret}
I'(\delta, \theta) = \frac{1}{2}\Big(I + \Big[\frac{1}{2}Q((1+\cos \delta) + (1- \cos \delta)\cos 4 \theta) + \frac{1}{2}U(1-\cos \delta)\sin 4 \theta - V \sin \delta \sin 2 \theta \Big] \Big) \,.
\end{equation}
From Eq.~\ref{rotret} it is clear that circular and linear polarization are
modulated in different channels (2$\theta$ and 4$\theta$, respectively) that are
strictly orthogonal regardless of the calibration error, and that the modulation
efficiency is maximized for Stokes $V$ when the retarder is a quarter-wave plate 
(i.e.\ when $\delta$=$\pi/2$). 
Note that linear polarization is then still modulated with half the efficiency.
We implement this modulation approach as a spatial modulation with a patterned
quarter-wave retarder that is co-located with the spectrograph slit.
The corresponding modulated spectra for 100\% $Q$, $U$, and $V$ are presented 
in Fig.~\ref{fig:polmod1}, which demonstrates the individual channels for 
linear and circular polarization.
We choose to implement this method in combination with 
a polarizing beam-splitter, that allows for the mitigation of the intensity 
imbalance in Stokes $Q$ (see Eq.~\ref{rotret} and the left and right top panels 
in Fig.~\ref{fig:polmod1}), and, more importantly, issues that arise when the 
slit is not homogeneously illuminated and/or not uniform in 
width\cite{Bagnulo+2009}.

The required patterned quarter-wave plate is readily produced using 
state-of-the-art liquid-crystal techniques. A direct-write 
technique\cite{MiskiewiczEscuti2014} is used to impose any fast-axis orientation 
pattern onto any optical substrate, with a resolution down to $\sim$1 $\mu$m.
Next, several layers of liquid crystals with different properties (type of
molecule, layer thickness, twist) can be combined to yield close-to-quarter-wave 
retardance over a large spectral bandwidth (e.g.~the entire visible 
range)\cite{Komanduri+2013}.
Such a quarter-wave plate resembles a so-called ``polarization 
grating''\cite{OhEscuti2008,Packham+2010}, that very effectively splits circular 
polarization into diffraction orders $\pm$1 when the retardance is half-wave, 
and the geometric phase\cite{Escuti+2016} dominates the optical propagation.
In our case, we apply large ``grating periods'' and half the efficiency 
(because of the quarter-wave retardance) such that the diffraction effects 
are under control. 

\begin{figure} [t]
   \begin{center}
   \includegraphics[width=0.9\textwidth]{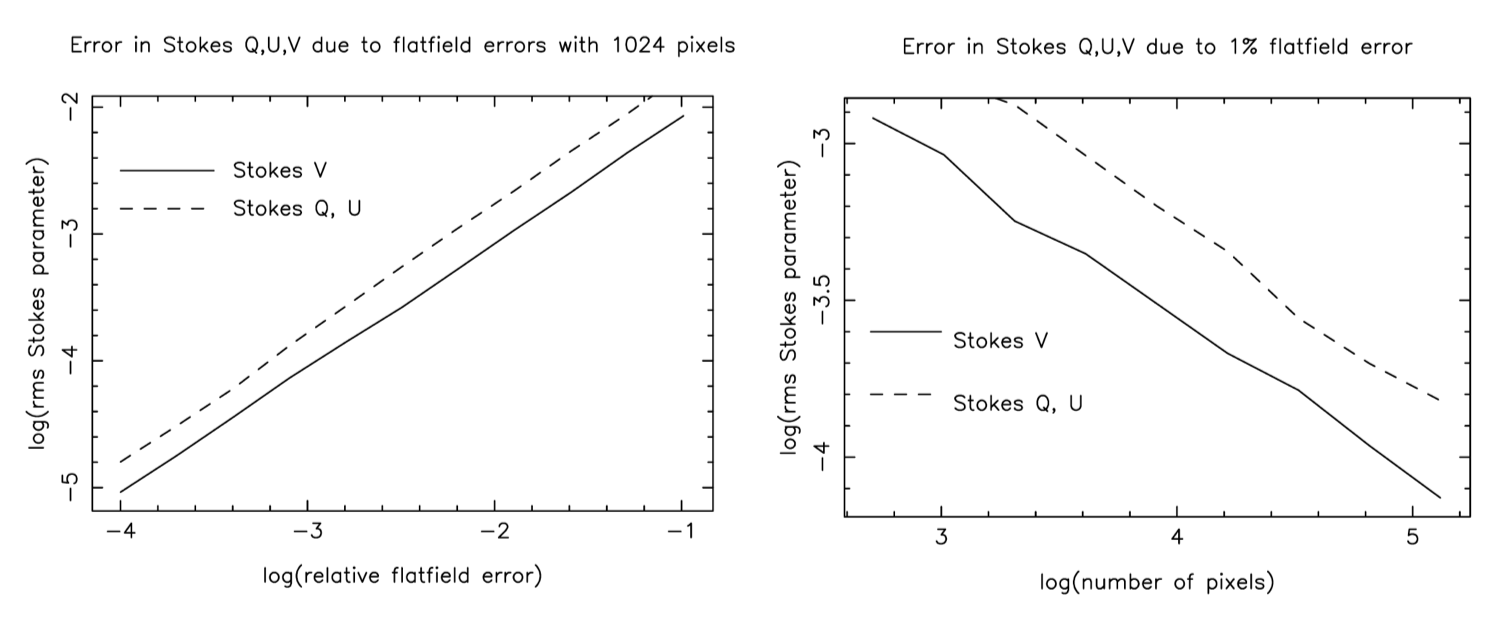}
   \end{center}
   \caption{ \label{fig:flatfield} 
            Simulations of spurious polarization effects due to limited 
            flat-field calibration (pixel-to-pixel variations).}
\end{figure} 

\begin{figure} [p]
       \begin{center}
   \includegraphics[width=0.85\textwidth]{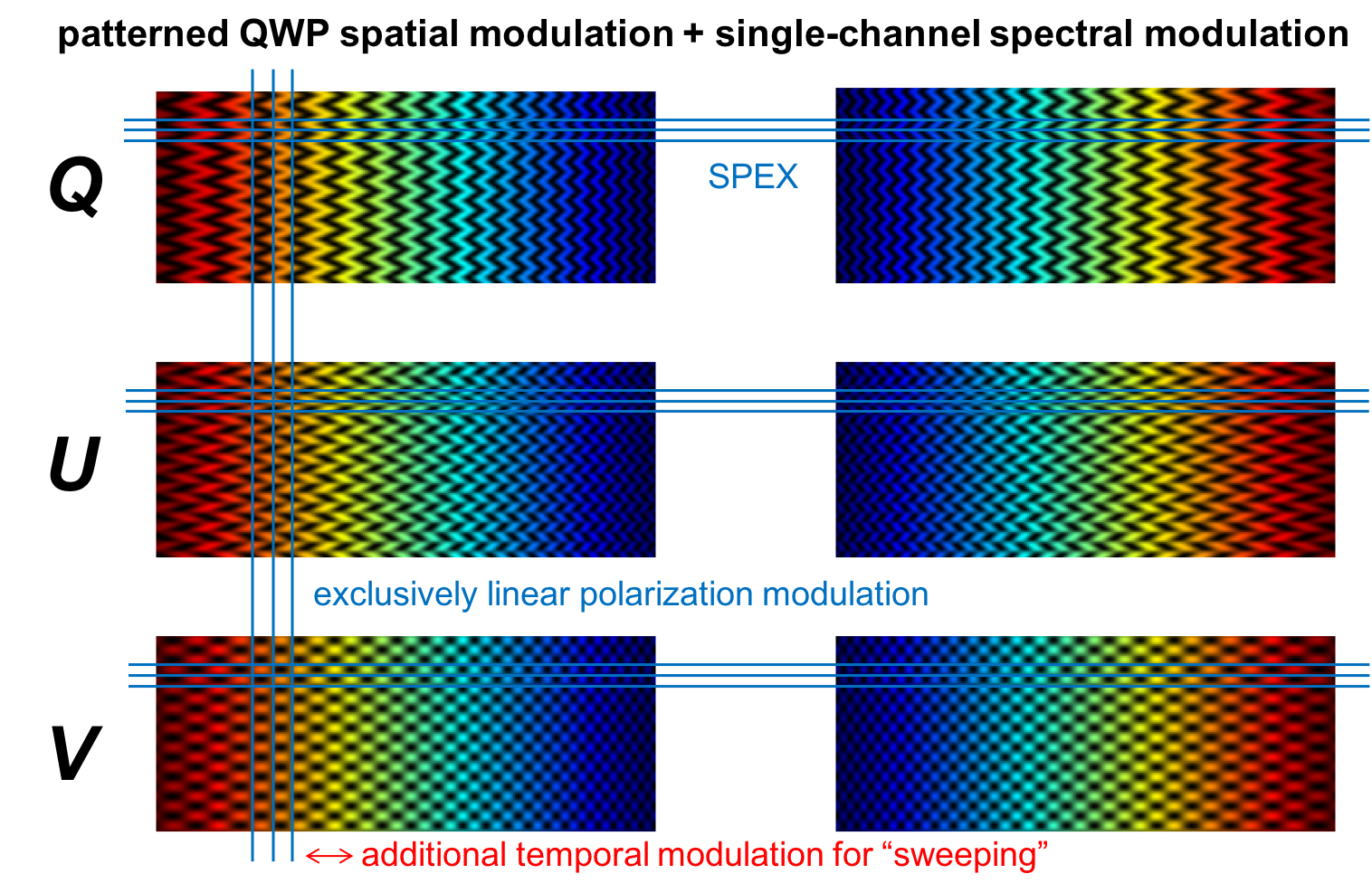}
   \end{center}
   \caption{ \label{fig:polmod2} 
            Modulation patterns for spatial modulation with a patterned 
            quarter-wave retarder in combination with a {\em single} 
            spectral modulator crystal. }
\end{figure} 
\begin{figure} [p]
   \begin{center}
   \includegraphics[width=0.85\textwidth]{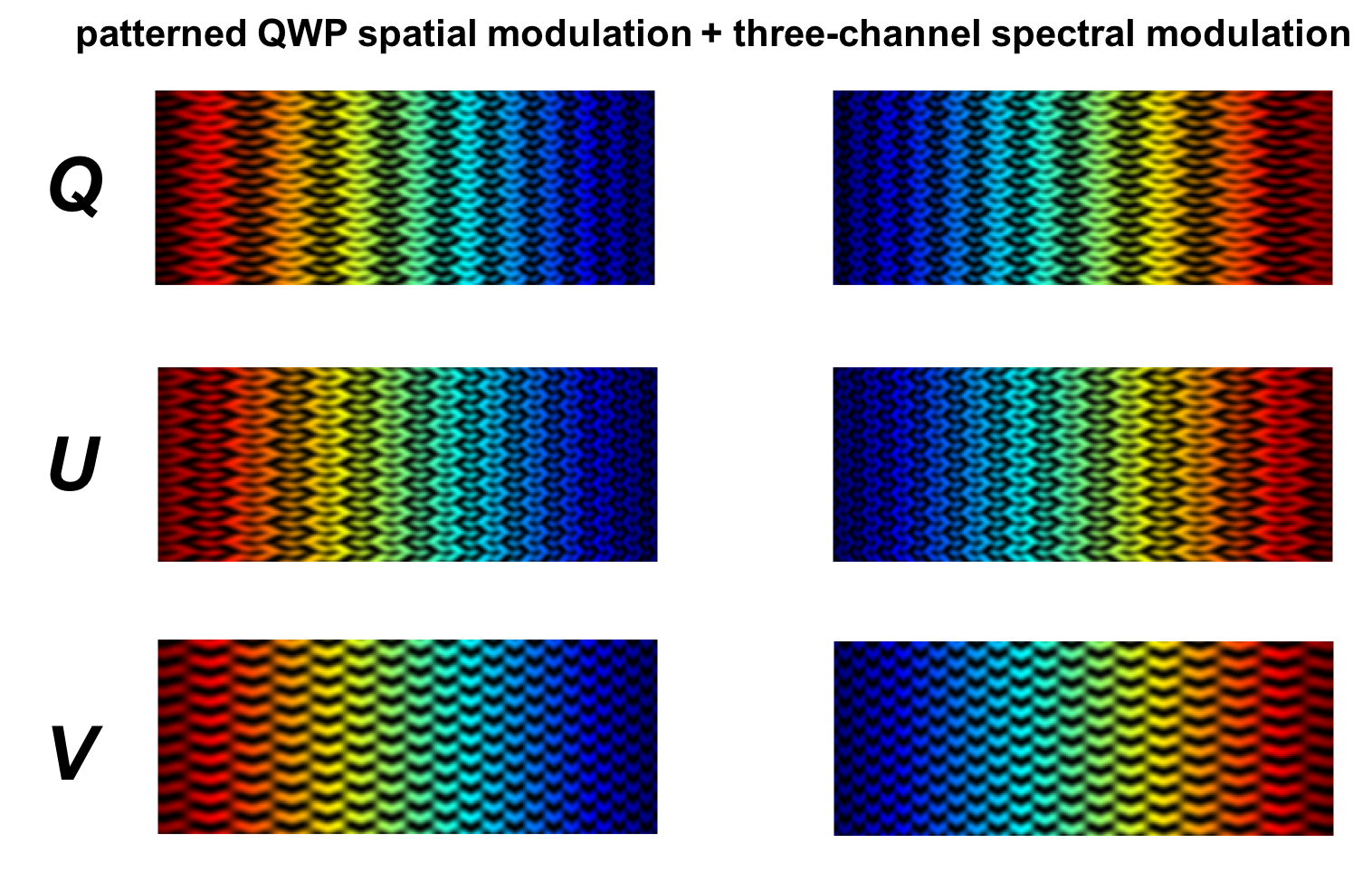}
   \end{center}
   \caption{ \label{fig:polmod3} 
            Modulation patterns for spatial modulation with a patterned 
            quarter-wave retarder in combination with {\em two} 
            spectral modulator crystals.}
\end{figure} 

After the spatial modulation and the dual-beam implementation, the most 
important source of systematic error to the polarimetry stems from limited
calibration accuracy of the pixel-to-pixel responses (the ``flat-field'').
Figure~\ref{fig:flatfield} shows how this error gets averaged down by 
spreading the modulation over many pixels in the cross-spectral direction.
A flat-field calibration accuracy of $\sim$1--0.1\% is deemed feasible, which corresponds to a realistic amount of pixels ($\sim$1000) to achieve $\sim$10$^{-4}$ polarimetric sensitivity for Stokes $V$.

To further mitigate such systematics, we experimented with an additional 
layer of spectral modulation. 
Figure~\ref{fig:polmod2} shows the resulting modulation patterns when a 
single thick retarder crystal is added as a spectral polarization modulator behind the patterned quarter-wave retarder.
It is clear that $Q$, $U$, and $V$ obtain an even more different appearance, 
although the modulation efficiency in linear polarization has increased.
Interestingly, specific rows and columns in the modulation patterns exhibit 
exclusively modulation in linear polarization, that can as such be used 
to assess systematics. 
The rows in which the quarter-wave retarder is oriented at $\pm$45$^\circ$ 
have a fully efficient spectral modulation for linear polarization\cite{Snik+2009}, 
for which advanced data-analysis routines have been developed in the context of the 
SPEX instrument suite\cite{Snik+2010,vanHarten+2014,Rietjens+2019}.
Also, specific columns for which the spectral modulator has a retardance modulo 
$\pm\pi/2$, there is no modulation for $V$.
As linear polarization will typically be observed for a wide range of 
polarization angles, all these pixels can be used to assess the effects of 
limited flat-field calibration on the final averaged circular polarization 
spectrum of Earth.
One more layer of temporal modulation could be added to the spectral modulator
(e.g.~an LCVR\cite{GarciaParejo:19}) to sweep these pixels over the entire 
spectrum.
In Fig.~\ref{fig:polmod3}, we show the effect of a full-Stokes spectral 
modulation\cite{OkaKato1999} that yields even more pronounced modulation patterns.
Interestingly and incidentally, Stokes $V$ appears as a pattern of little 
'V's.


\begin{figure} [!b]
   \begin{center}
   \includegraphics[width=\textwidth]{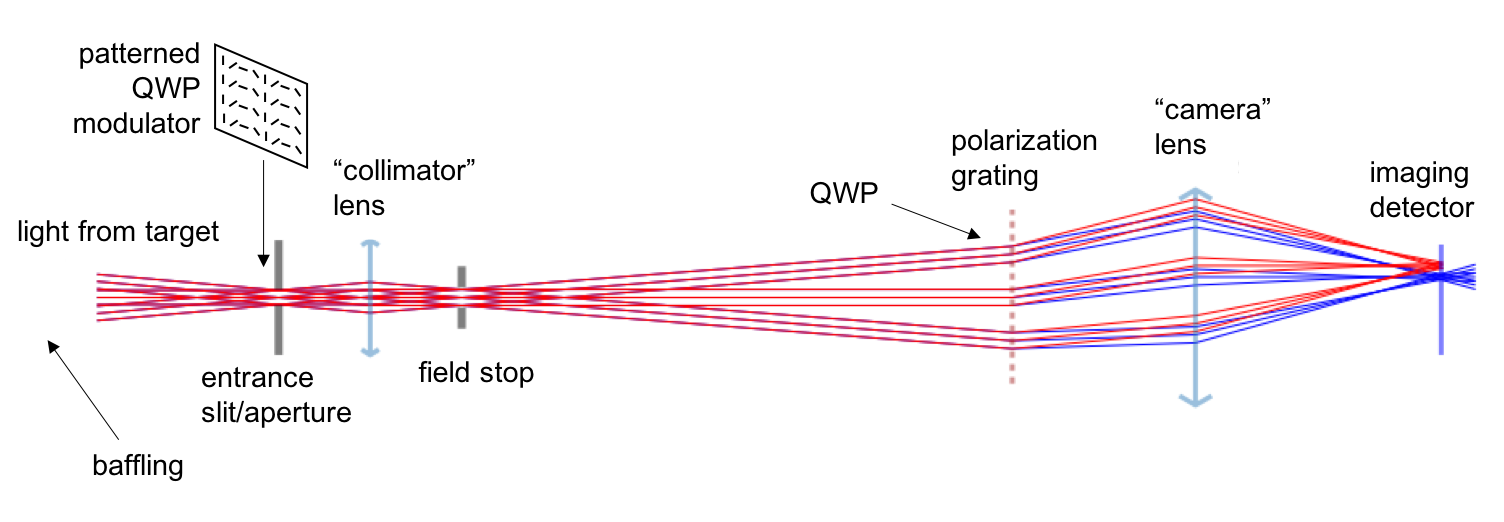}
   \end{center}
   \caption{ \label{fig:optdes} 
            Sketch of the optical design for LSDpol. Only the +1 diffraction order of the polarization grating is traced.}
\end{figure} 

\section{OPTICAL DESIGN}

The adopted modulation approach drives the optical design of the spectrograph.
Firstly, the spatial modulator needs to be co-located with the spectrograph slit.
Secondly, the modulator and slit need to be illuminated uniformly.
And thirdly, the polarization modulator should be the very first element 
of the optical train, to minimize the effects of cross-talk due to any 
fore-optics\cite{Kemp+1987}.
This means that the slit acts both as the spectrograph slit, as well as the 
entrance aperture of the instrument, a veritable \emph{contradictio in terminis}!
To keep the spectrograph design simple, we still adopt the configuration of
collimator--grating--camera, but this time the ``collimator'' produces an 
image of the target at its back focal point.
This is where we locate a field stop, that, together with baffling in front 
of the slit, defines the field of view over which the instrument integrates for a
snapshot observation.
The rays for different points along the slit hit the grating under different 
angles, but it turns out that for the medium spectral resolution required for the 
simple modulation approach as shown in Fig.~\ref{fig:polmod1}, this is acceptable.
We implement an off-the-shelf polarization grating\cite{OhEscuti2008,Packham+2010}\footnote{\href{https://www.edmundoptics.com/f/polarization-gratings/39543/}{https://www.edmundoptics.com/f/polarization-gratings/39543/}}, that together with an achromatic quarter-wave plate in front, combines linear polarization beam-splitting with high-efficiency broadband diffraction in spectral orders $\pm$1.
A ``camera'' lens then reimages the slit onto the detector and produces the 
two complementary spectra.
The ensuing optical design is sketched in Fig.~\ref{fig:optdes}.

As the slit determines both the spectral resolution as well as the photon flux,
there is an interesting trade-off for the slit width, which for now we fix at 100
$\mu$m for a tentative design for the ISS altitude and ground speed. 
The slit needs to be as long as possible to catch as many photons as possible, and  to average out systematic effects along the spatial polarization modulation direction.
Both the slit as well as the spatial polarization modulator (acting as an 
inefficient polarization grating) diffract light, in perpendicular directions.
In combination with the field stop, this leads to some vignetting (which also the
baffling will induce).
A careful analysis and a context camera will therefore be required to establish 
the exact footprint on Earth pertaining to the signal for every spectral 
recording.


\section{CONCLUSIONS \& OUTLOOK}

The challenging requirements of Sect.~\ref{req} have led us to the introduction 
of a novel spatial modulation approach in combination with a very unconventional 
spectrograph optical design (although no hallucinogens were necessary to come up with the LSDpol concept; the imagery in Figs.~\ref{fig:polmod1}--\ref{fig:polmod3} was sufficient).
We are currently building a prototype instrument cf.~the design in Fig.~\ref{fig:optdes}, which will first be calibrated and tested in the lab.
Next, we will deploy it in field and aircraft campaigns, and compare the results with measurements from TreePol 1\&2.
We will further develop and miniaturize the LSDpol instrument concept based on our experiences, and work in the context of the MERMOZ project\footnote{\href{http://www.saintex.unibe.ch/research/research\_interests/}{http://www.saintex.unibe.ch/research/research\_interests/}} towards an implementation for ISS/Bartolomeo (or, alternatively, a cubesat), which is the ideal platform to create a crucial map of the circular polarization signals for the only planet we currently know harbors life.
This instrument is ideally flown in conjunction with other dedicated Earth-as-an-exoplanet instrumentation, such as LOUPE.
Furthermore, the linear spectropolarimetry of LSDpol is ideally cross-validated with the SPEXone instrument that is being constructed for the PACE mission\cite{Rietjens+2019} to provide essential measurements of the influence of atmospheric aerosols on our climate and our health.
As such, MERMOZ/LSDpol can also contribute to Earth science, in addition to detecting life on Earth to enable the detection of life elsewhere in the universe.

\acknowledgments 
The research of FS and DSD leading to these results has received funding from the European Research Council under ERC Starting Grant agreement 678194 (FALCONER). 
This contribution of HJH has received funding from the European Research Council (ERC) under the European Union’s Horizon 2020 research and innovation programme
(projects Four Aces and EXOKLEIN with grant agreement numbers 724427 and 771620, respectively). JK, HJH, AP \& B-OD are acknowledging support from the NCCR PlanetS, the Swiss National Science Foundation and the University of Bern's Centre for Space and Habitability (CSH) that are funding the feasibility study of the MERMOZ project.

\bibliography{LSDpolSPIE2019} 
\bibliographystyle{spiebib} 

\end{document}